\documentclass[aps,prl,reprint,superscriptaddress]{revtex4-1}

\usepackage{graphicx}
\usepackage{dcolumn}
\usepackage{bm}
\usepackage{lineno}
\usepackage{soul}
\usepackage{color}
\usepackage{array}
\usepackage{booktabs}
\usepackage{multirow, makecell}

\begin{document}


\title{Electronic Spin transition in FeO$_{2}$: evidence for Fe(II) with peroxide O$_{2}^{2-}$}



\author{Bo Gyu Jang}
\affiliation{Department of Chemistry, Pohang University of Science and Technology, Pohang 37673, Korea}

\author{Jin Liu}
\affiliation{Department of Geological Sciences, Stanford University, Stanford, California 94305, USA}

\author{Qingyang Hu}
\affiliation{Department of Geological Sciences, Stanford University, Stanford, California 94305, USA}
\affiliation{Center for High Pressure Science and Technology Advanced Research (HPSTAR), Shanghai 201203, China}

\author{Kristjan Haule}
\affiliation{Department of Physics and Astronomy, Rutgers University, Piscataway, New Jersey 08854, USA}

\author{Ho-Kwang Mao}
\affiliation{Center for High Pressure Science and Technology Advanced Research (HPSTAR), Shanghai 201203, China}

\author{Wendy. L. Mao}
\affiliation{Department of Geological Sciences, Stanford University, Stanford, California 94305, USA}
\affiliation{Stanford Institute for Materials and Energy Sciences, SLAC National Accelerator Laboratory, Menlo Park, California 94025, USA}

\author{Duck Young Kim}
\email{duckyoung.kim@hpstar.ac.cn}
\affiliation{Center for High Pressure Science and Technology Advanced Research (HPSTAR), Shanghai 201203, China}

\author{Ji Hoon Shim}
\email{jhshim@postech.ac.kr}
\affiliation{Department of Chemistry, Pohang University of Science and Technology, Pohang 37673, Korea}
\affiliation{Department of Physics and Division of Advanced Nuclear Engineering, Pohang University of Science and Technology, Pohang 37673, Korea}


\date{\today}

\begin{abstract}
The discovery of FeO$_{2}$ containing more oxygen than hematite (Fe$_{2}$O$_{3}$) that was previously believed to be the most oxygen rich iron compounds, has important implications on the study of the deep lower mantle compositions. Compared to other iron compounds, there are limited reports on FeO$_{2}$ making studies of its physical properties of great interest in fundamental condensed matter physics and geoscience. Even the oxidation state of Fe in FeO$_{2}$ is the subject of debate in theoretical works and there have not been reports from experimental electronic and magnetic properties measurements. Here, we report the pressure-induced spin state transition from synchrotron experiments and our computational results explain the underlying mechanism. Using density functional theory and dynamical mean field theory, we calculated spin states of Fe with volume and Hubbard interaction $U$ change, which clearly demonstrate that Fe in FeO$_{2}$ consists of Fe(II) and peroxide O$_{2}^{2-}$. Our study suggests that localized nature of both Fe 3$d$ orbitals and O$_{2}$ molecular orbitals should be correctly treated for unveiling the structural and electronic properties of FeO$_{2}$.

\end{abstract}


\maketitle


\clearpage
\newpage
\pagebreak

The recent discovery of pyrite-structured FeO$_{2}$ at high pressures and temperatures has generated significant interest as an alternative explanation for seismic observations in Earth's deep mantle and for its potential implications on volatile storage and cycling within our planet.\cite{Hu2016, Hu2017, Liu2017, Nishi2017}. To date, the electronic and magnetic properties of FeO$_{2}$ under high pressure are still poorly understood. Also there have been controversy surrounding the oxidation state of FeO$_{2}$ under high pressure. It was suggested that the existence of peroxide, O$_{2}^{2-}$ in FeO$_{2}$ induces the oxidation state of Fe(II) similar to FeS$_{2}$. A band type insulator-to-metal transition is expected to produce Fe$^{(2+\delta)+}$ by the elongation of the O$_{2}$ dimer bond length\cite{Jang2017}. In another theoretical work by S. S. Streltsov \textit{et al.}, however, Fe(III) with O$_{2}^{3-}$ state was suggested due to the long O$_{2}$ dimer bond length compared to the usual bond length of peroxide\cite{Streltsov2017}. 

The chemical stability of FeO$_{2}$ and FeO$_{2}$H has also debated. Q. Hu \textit{et al.} reported that FeO$_{2}$H$_{x }$ (0$\leq x \leq$1) can be synthesized under lower mantle condition and $x$ can vary depending on the external condition by releasing H$_{2}$ molecule\cite {Hu2016, Hu2017, Liu2017}. However, M. Nishi \textit{et al.} claimed that FeO$_{2}$H is much more stable than FeO$_{2}$ with H$_{2}$ based on DFT calculations\cite {Nishi2017}. The dehydrogenation of FeO$_{2}$H is nothing but the oxidation process of FeO$_{2}$H so that this issue is also closely related to the oxidation state of Fe and the O$_{2}$ dimer. The discrepancy in the stability of FeO$_{2}$H may come from the description of the oxidation state in DFT calculation which is simliar with the first issue.

Transition metal oxides (TMO) like iron oxides exhibit rich phase diagrams, which is mainly originated from the electron correlation effects of the 3$d$ orbitals. Many TMO such as FeO, Fe$_{2}$O$_{3}$, and MnO show Mott-type insulating behavior at ambient pressure. This Mott insulating state can be broken down under high pressure, where the $U/W$ ($U$: Coulomb interaction, $W$: band width) ratio becomes smaller and it eventually reaches metallic states \cite{Imada1998}. Usually this metal-insulator transition (MIT) is accompanied by a volume-collapsing spin state transition (SST), from a high spin (HS) Mott insulator to low spin (LS) state \cite{Cohen1997, Kunes2008, Kunes2009, Ohta2012, Leonov2015}.

The crystal structure of FeO$_{2}$ and FeS$_{2}$ is similar to that of B1 type FeO. This Pa-3 pyrite-type structure can be easily obtained by replacing oxygen atoms in B1 type FeO with X$_{2}$ (X=O or S) dimers. Despite of high similarity in the crystal structure and the same oxidation state of Fe in FeO, FeS, and FeS$_{2}$, which are well studied compounds, they show different behavior under pressure. FeO and FeS undergo a SST (S=2 to S=0) accompanied with breakdown of Mott insulating phase with partially filled $t_{2g}$ and $e_{g}$ orbitals under high pressure\cite{Badro1999, Rueff1999, Ohta2012, Leonov2015, Ushakov2017}   . On the other hand, FeS$_{2}$ does not display Mott type MIT and SST under high pressure. It is reported to be a non-magnetic (S=0) band insulator with fully occupied $t_{2g}$ orbitals \cite{Rueff1999, Fujimori1996, Miyahara1968, Chattopadhyay1985}. A potential SST in FeO$_{2}$ under pressure is not only an interesting issue in and of itself but also is very relevant to resolving the controversy surrounding  the oxidation state of Fe in this compound.

In this paper, we provide evidence for a SST in FeO$_{2}$ under high pressure by means of synchrotron experiments and first-principles calculations. From this study, we find that the SST in FeO$_{2}$ observed in our experiment originates from Fe (II) state rather than Fe (III) state. The delocalization error of DFT may have caused the controversy in the oxidation state and the chemical stability between previous studies\cite{Jang2017, Streltsov2017, Hu2016, Hu2017, Liu2017, Nishi2017}. However, the oxidation state of Fe can be (2+$\delta$)+ depending on external conditions, specially under high pressure due to short Fe-O bonding.

\section{Results}

\begin{figure} 
\includegraphics[width=\linewidth]{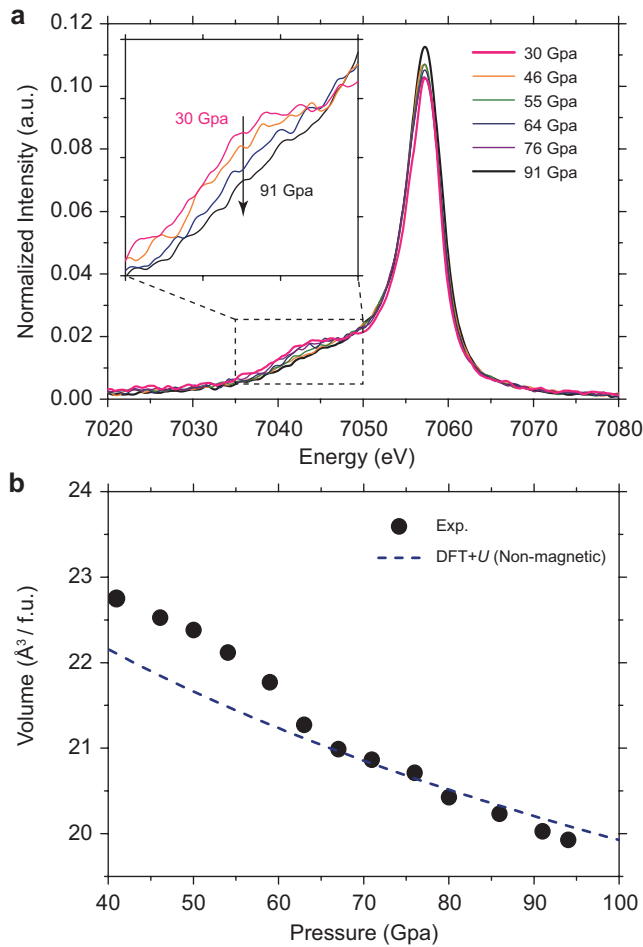}
\caption{\textbf{a} X-ray emission spectra for FeO$_{2}$ as a function of pressure. The satellite peak at around 7040 eV is characteristic of the high-spin state. \textbf{b} Pressure-dependent FeO$_{2}$ volume. Blue dashed line are calculated from non-magnetic calculation with DFT+$U$ method ($U$=5 eV). 
}
\label{fig:Fig1}
\end{figure}

Figure 1\textbf{a} displays the X-ray emission spectroscopy (XES) measurements on FeO$_{2}$ under high-pressure. The typical Fe K $\beta$ main emission peak is located around 7058 eV. The most important feature is a satellite peak located near 7040 eV which is a typical K $\beta'$ peak of Fe \cite{Badro1999, Rueff1999}. This K $\beta'$ peak collapses with increasing pressure and this is a typical evidence of SST of iron oxides. The unit cell volume of FeO$_{2}$ was determined from X-ray diffraction (XRD) patterns between 40 and 94 GPa. A volume collapse of approximately 6\% was observed between 50 and 65 GPa, where FeO$_{2}$ undergoes a SST of Fe (Fig. 1\textbf{b}). The pressure-volume relation of FeO$_{2}$ has been calculated from non-magnetic (NM) DFT+$U$ method. It is in good agreement with experimental points at pressures above 65 GPa, where FeO$_{2}$ is in the LS state as shown in Fig. 1\textbf{b}. Therefore, these results indicate that FeO$_{2}$ undergoes an HS to LS (S=0) transition from 50 to 65 GPa. This NM behavior in LS state implies the Fe(II) $d^{6}$ nature of FeO$_{2}$.

\begin{figure*}
\includegraphics[width=0.77\linewidth]{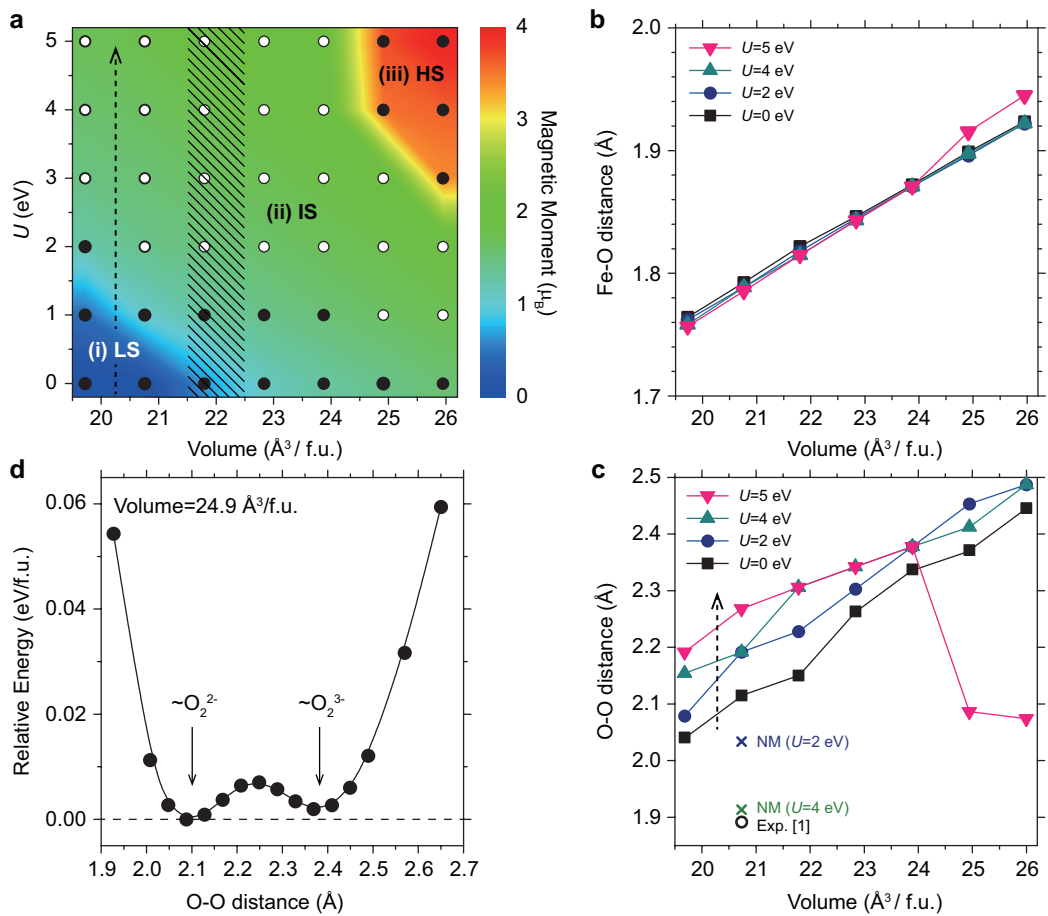}
\caption{\textbf{a} Calculated volume ($V$)-Coulomb interaction ($U$) phase diagram of FeO$_{2}$. Colors present the size of magnetic moment. Open and closed circles indicate the insulating and metallic phases respectively. Dashed region indicates the experimentally observed transition point. Calculated \textbf{b} Fe-O bond length and \textbf{c} O$_{2}$ dimer bond length depending on the $U$ value. The arrow in \textbf{a} and \textbf{c} indicate that the magnetic moment increases as the O$_{2}$ dimer bond length increases. There are abrupt change both in Fe-O and O-O bond length at $U$=5 eV. \textbf{d} Relative energy depending on the O$_{2}$ dimer bond length at volume of 24.9 {\AA}$^{3}$/f.u. with $U$=5 eV. There are two local minima depending on the O$_{2}$ dimer bond length.
}
\label{fig:Fig2}
\end{figure*}

We performed spin polarized calculations using DFT+$U$ method to investigate the SST observed in high-pressure XES and XRD experiments aforementioned. The internal parameters are fully relaxed with given volume and the Coulomb-interaction $U$ value while keeping the Pa-3 symmetry. Figure 2\textbf{a} shows the calculated volume ($V$) - Coulomb interaction ($U$) phase diagram of FeO$_{2}$. The color indicates the size of magnetic moment while open and closed circles indicate insulating and metallic phase, respectively. Black dashed region indicates the experimentally observed transition point. Three phases are clearly distinguished by the size of magnetic moment: (i) NM LS region (Blue), (ii) intermediate spin state IS region (Green), and (iii) HS (S$\sim$2) region (Red). The IS state is believed to be unusual for Fe (II) oxidation state. The size of the magnetic moment changes abruptly between phase (ii) and (iii), while it increases rather smoothly between phase (i) and (ii). (See Supplementary Fig. 1\textbf{c}) This phase diagram itself looks reasonable, because the size of magnetic moment increases as the volume and $U$ value increases. However, it cannot describe the SST observed in the experiment due to the unusual IS state, region (ii) as following reason.

At small $U$ value ($U$ $<$ 2 eV), the transition between LS to IS is observed but it is not accompanied with the abrupt change in volume and magnetic moment. It shows continuous growth in the magnetic moment which is not a first-order transition. Also the energy-volume curves do not show the first-order transition behavior. At $U$=0 eV, the energy-volume curve of IS gradually diverges from that of LS as the magnetic moment gradually increases. (See Supplementary Fig. 1.) Although the HS (S$\sim$2) state can be observed at large $U$ value ($U$ $\geq$  2 eV), the intermediate spin state can be stabilized even at the smallest volume we tested. Also the calculated transition volume is much larger than the experimentally observed transition volume.

\begin{table} [!b]
\begin{ruledtabular}
\begin{center}
\centering
\caption{Ionic radius of Fe atom depending on its spin and oxidation state\cite{Shannon1976}.}
\renewcommand{\arraystretch}{1.2}
\begin{tabular}{>{\centering}m{.2\linewidth}|>{\centering} m{.4\linewidth} | >{\centering}m {.4\linewidth}}
              & Fe(II) $d^{6}$ & Fe(III) $d^{5}$\tabularnewline \hline
LS &         0.75 {\AA}               &  0.69 {\AA}    \tabularnewline \hline
HS &         0.92 {\AA}               &  0.79 {\AA}    \tabularnewline
\end{tabular}
\end{center}
\end{ruledtabular}
\end{table}

To investigate the origin of this discrepancy, we analyze the bond length in detail. It is well known that the ionic radius of metal atoms varies according to their spin state and oxidation state. The ionic radius of the Fe atom depending on their spin and oxidation state are listed in Table I \cite{Shannon1976}. Also O$_{2}$ dimer bond length is related to its oxidation state. The bond length of a neutral O$_{2}$ molecule is 1.21 {\AA} and it increases as the O$_{2}$ dimer takes more electrons by occupying the anti-bonding states. For example, the bond length of superoxide (O$_{2}^{-}$) in KO$_{2}$ is 1.28 {\AA} \cite{Kanzig1976} and that of peroxide (O$_{2}^{2-}$) in BaO$_{2}$ is 1.49 {\AA} \cite{Wong1994}. A quantitative approach on the bond length can offer an effective way of understanding the calculated phase diagram in Fig. 2. 

\begin{figure*}
\includegraphics[width=\linewidth]{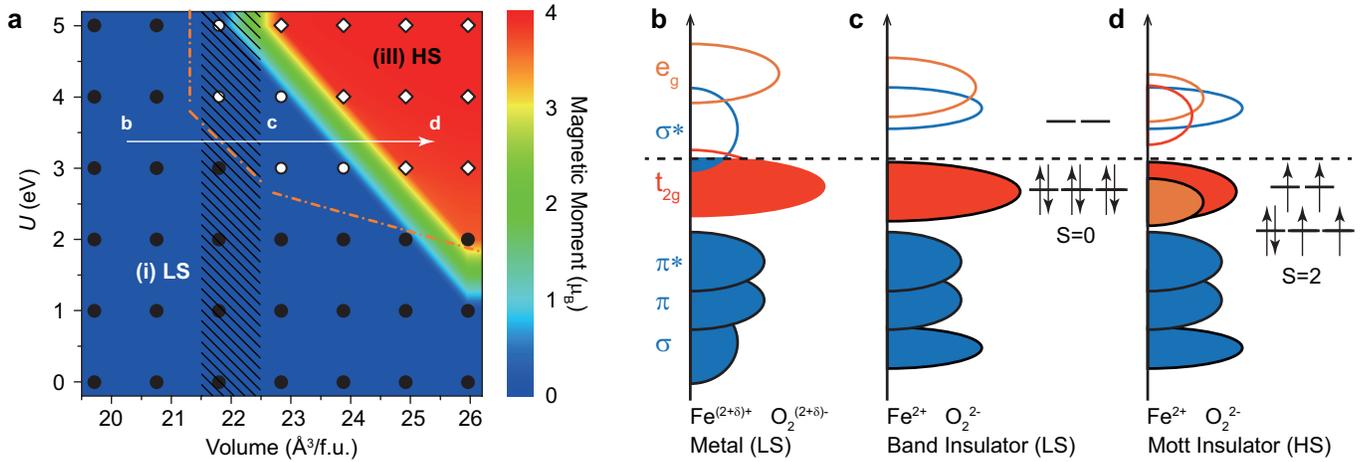}
\caption{\textbf{a} Calculated $V-U$ phase diagram of FeO$_{2}$ using fixed O$_{2}$ dimer bond length of 1.8 {\AA}. Open diamonds indicate the Mott insulating state. Orange dash-dot line indicates the boundary for Fe (II) with peroxide state. LS region is divided into \textbf{b} and \textbf{c} regions by the orange dash-dot line. The electronic structure evolves from \textbf{b} LS metalic state to \textbf{c} LS band insulator, and finally to \textbf{d} HS Mott insulator along the white guide line in \textbf{a}. 
}
\label{fig:Fig3}
\end{figure*}

Figure 2\textbf{b} and \textbf{c} show the bond length of Fe-O and O-O at several volumes and $U$ values. It should be noted that the O$_{2}$ dimer bond length increases and moves away from the experimental value as the $U$ value increases at small volume (Dashed arrow in Fig. 2\textbf{c}) and this trend is in contrast to the tendency observed in the NM calculation denoted by cross marks\cite {Jang2017}. In NM calculation, the O$_{2}$ dimer bond length is getting shorter and closer to the experimental value as the $U$ value increases as shown in Fig. 2\textbf{c} and Fig. 3 in Ref.\cite{Jang2017}. This results also support that the LS state of FeO$_{2}$ is non-magnetic. 

All the calculated values for the O$_{2}$ dimer bond length are longer than the well-known peroxide bond length of $\sim$1.49 {\AA} and also longer than 1.8 {\AA} which is the critical value for Fe(II) and the peroxide state reported in our previous study \cite {Jang2017}. As O$_{2}$ dimer bond length increases, the size of magnetic moment gradually increases to induce an unusual IS state with an Fe $^{(2+\delta)+}$ and O$_{2}^{(2+\delta)-}$ configuration. As the $U$ value and volume increase, the size of the magnetic moment of the IS state eventually converge to $\sim$2 $\mu_{B}$ which is close to the magnetic moment of Fe(III) LS compounds. Large $U$ values ($U$=3, 4 eV) with large volumes eventually produce the localized Fe $d$ orbital of S=5/2 state. 

Furthermore, an abrupt bond length change is observed at large volume with $U$=5 eV both in Fe-O and O-O bonding. It is due to the competition between the nearly Fe (III) HS (S=5/2) state with long O$_{2}$ dimer (short Fe-O bond) and Fe (II) HS (S=2) state with short O$_{2}$ dimer (long Fe-O bond). Figure 2\textbf{d} clearly shows two local minima depending on the O$_{2}$ dimer bond length at 24.9 {\AA}$^{3}$/f.u. with $U$=5 eV. The IS to HS transition at $U$=4 eV is close to [Fe (III), S=1/2] to [Fe (III), S=5/2] transition while the transition at $U$=5 eV is close to [Fe(III), S=1/2] to [Fe(II), S=2] state. The SST at $U$=5 eV is accompanied by the change in the oxidation state of Fe and O$_{2}$ dimer so that the abrupt bond length changes are observed as shown in Fig. 2\textbf{b-d}.

The analysis of bond length indicates that the discrepancy between the experiment and the calculation arises from the Fe$^{(2+\delta)+}$ and O$_{2}^{(2+\delta)-}$ state which makes unusual IS state for Fe (II) case. We perform a test calculation with a fixed short O$_{2}$ dimer bond length of 1.8 {\AA}. Surprisingly, the IS state region disappears and just LS (S=0) state and HS state (S=2) remain as shown in Fig. 3\textbf{a}. Orange dash-dot line indicates the boundary for Fe (II) with peroxide state. Below this line, it shows metallic behavior due to the small overlap between the Fe $t_{2g}$ band and O$_{2}$ dimer $\sigma$* band as shown in Fig. 3 (b). Above the line, FeO$_{2}$ has Fe$^{2+}$ with peroxide state. The SST occurs around at 22 {\AA}$^{3}$/f.u. with reasonable $U$ value (5 eV) in DFT+$U$ calculations. This phase diagram is in good agreement with the experimental result. The schematic electronic structures of each regions are described in Fig. 3\textbf{b-d}. This results imply that the SST of FeO$_{2}$ originates from Fe(II) with a peroxide state. However, the oxidation state of Fe can be (2+$\delta$)+ under high pressure due to the short Fe-O bonding as already discussed in our previous study\cite{Jang2017}.

This phenomenon is also confirmed by a DFT+DMFT calculation. With Fe $d$ orbital occupancy of 6, it also shows direct SST from S=2 to S=0 with short O$_{2}$ dimer bond length. When a long O$_{2}$ dimer from the structural relaxation is considered, the DFT+DMFT calculation also fails to reproduce the experimental HS-LS transition 

We calculate local spin susceptibility, $\chi_{loc}$ for different bond length at volume of 24.9 {\AA}$^{3}$/f.u. which is clearly larger than the experimental transition volume (See Supplementary Fig. 2). With an O$_{2}$ dimer bond length of 1.49 {\AA} (long Fe-O bond), it shows HS state. The calculated $\chi_{loc}$ at several temperature points are well fitted by the Curie-Weiss formula $\chi_{loc}$ =$C/T$, as expected for a local spin system. On the other hand, with an O$_{2}$ dimer bond length of 2.1 {\AA} (short Fe-O bond), which is obtained from DFT+$U$ structural relaxation, the calculated $\chi_{loc}$ are very small and almost $T$-independent due to NM solution. These DFT+DMFT results again suggest that the HS-LS transition of FeO$_{2}$ can occur only with Fe (II) and a peroxide state and is very sensitive to the O$_{2}$ dimer bond length (Fe-O bond length).

\begin{figure}
\includegraphics[width=\linewidth]{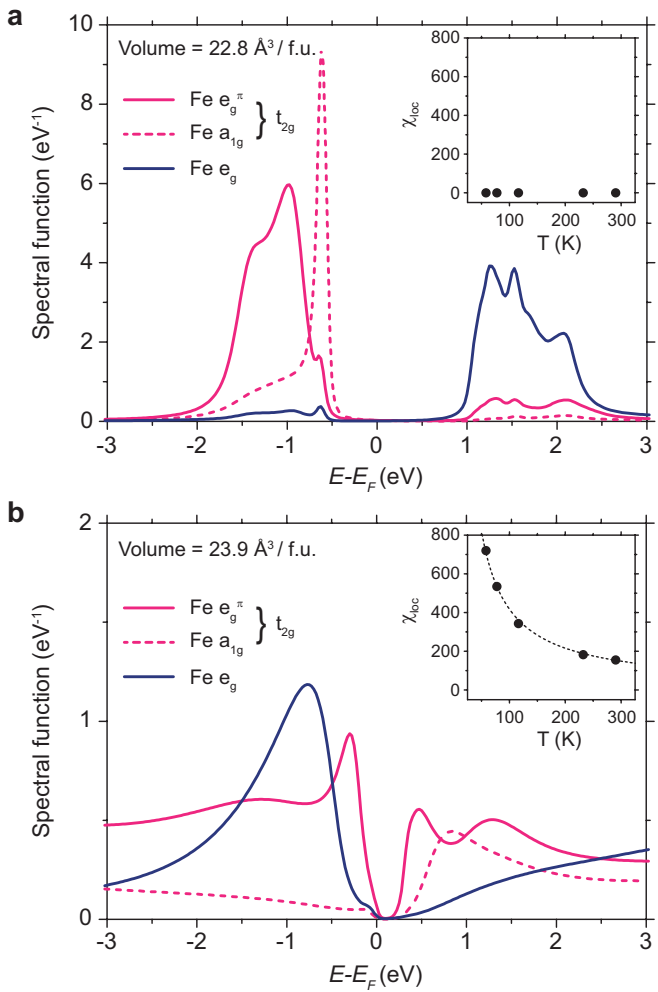}
\caption{Density of states (DOS) calculated from DFT+DMFT at volume \textbf{a} 22.8 {\AA}$^{3}$/f.u. (Band insulator, S=0) and  \textbf{b} 23.9 {\AA}$^{3}$/f.u. (Mott insulator, S=2). Inset shows the calculated temperature dependence of the local spin susceptibility at each volume. 
}
\label{fig:Fig4}
\end{figure}

The SST of FeO$_{2}$ is also accompanied with a Mott type transition. The open circles and diamonds in Fig. 3\textbf{a} indicate band insulator and Mott insulator state, respectively. In Fig. 4, the DFT+DMFT spectral function at 22.8 {\AA}$^{3}$/f.u. clearly show fully occupied $t_{2g}$ ($e_{g}^{\pi}$ and $a_{1g}$) and fully unoccupied $e_{g}$ band with band gap ($\sim$1 eV), which corresponds to the LS state of Fe(II). At larger volume of 23.9 {\AA}$^{3}$/f.u., it shows partially occupied $t_{2g}$ and $e_{g}$ band indicating HS state of Fe(II) with Mott insulating behavior. Inset  shows the calculated $\chi_{loc}$  for each volume. The calculated $\chi_{loc}$  at 22.8 {\AA}$^{3}$/f.u. shows $T$-independent NM behavior while the $\chi_{loc}$ at 23.9 {\AA}$^{3}$/f.u. follows the Curie-Weiss law due to its local spin moment.

\section{Discussion}
As mentioned above, the spin polarized DFT+$U$ calculation always gives a long O$_{2}$ dimer bond, which fails to explain the experimental result. There are several possible explanations for this discrepancy. The simplest is that the magnetic ordering configuration used in our calculations may be wrong. However, this can be easily excluded because if there is another energetically favorable magnetic configuration, the magnetic moment persists well below the 22 {\AA}$^{3}$/f.u., which is the experimental HS-LS transition boundary. Within magnetic ordering used in our calculations, the magnetic moment is already stabilized down to very small volumes compared to the experimental result as shown in Fig. 2\textbf{a}.

The other possible explanation is that there can be a structural phase transition accompanied with the SST. However, there is no experimental evidence supporting a structural phase transition. X-ray diffraction patterns in previous studies have not shown any abrupt change between 50 and 65 GPa, where the SST occurs\cite {Hu2017}. We also performed phonon calculations on FeO$_{2}$ and it does not show any dynamic instability. However, this may be an important issue for future study.

The remaining possibility for the discrepancy may come from the limitation of current calculation method. It is well-known that simple LDA/GGA underestimates the band gap which is connected with failure of description of molecular orbital energy level. It is due to the delocalization error of simple LDA/GGA. DFT calculation may underestimate the splitting between O$_{2}$ $\pi$*and $\sigma$* band, so O$_{2}$ dimer can easily get more electrons from Fe, making Fe$^{(2+\delta)+}$ and O$_{2}^{(2+\delta)-}$ states. In other words, the calculations easily predict FeO$_{2}$ as a metallic system and prefer the structure with long O$_{2}$ dimer. S. S. Streltsov \textit{et al} \cite{Streltsov2017}. interpreted the oxidation state of Fe in FeO$_{2}$ as 3+ rather than 2+ due to this problem. A similar issue was reported in NiSe$_{2}$ which also adopts a pyrite-type structure with Se$_{2}$ dimers in it. The simple DFT calculation overestimates the Se$_{2}$ dimer bond length by $\sim$0.15 {\AA} compared to the experimental value \cite{Kunes2010, Moon2015}. The controversy on the oxidation state of Fe may be due to this reason. 

In addition, if FeO$_{2}$ comes close to having Fe$^{3+}$ and O$_{2}^{3-}$, the $\sigma$* band becomes half-filled and the correlation effect of $\sigma$* orbital should be considered like KO$_{2}$ which has partially filled $\pi$* orbitals \cite{Kim2010}. Although we circumvent this problem by simply adopting the short O$_{2}$ dimer in the structure, this is an important issue for future research. A further study with a more precise description on the localized picture of O$_{2}$ molecular orbitals is therefore recommended.
 

We investigate the spin state transition of FeO$_{2}$ by using both experimental and first-principles approach. The abrupt volume collapsing occurs around 50-60 GPa and the X-ray emission spectrum indicate the existence of the SST. The typical K $\beta^{\prime}$  peak of Fe due to the HS state collapses with increasing pressure. However, the phase diagram obtained from the relaxed structure with DFT+$U$ calculation fails to describe the experimental SST due to the existence of an unusual IS state. We found that this IS states come from Fe$^{(2+\delta)+}$ and O$_{2}^{(2+\delta)-}$ state which originates from long O$_{2}$ dimer bond length (short Fe-O distance). The DFT+$U$ and DFT+DMFT calculations on the structures with short O$_{2}$ dimers give a correct description of the SST observed in the experiment. It suggests that the HS-LS transition of FeO$_{2}$ occurs with Fe (II) and a peroxide state and is very sensitive to the O$_{2}$ dimer bond length. We suggest that the precise description of not only Fe $d$ orbitals but also O$_{2}$ dimer molecular orbital should be considered in the future study to describe correct structural and electronic propertied of FeO$_{2}$. A detailed crystal structure study including the analyze of O$_{2}$ dimer bond length under pressure would also be very interesting to understand the physics of FeO$_{2}$. 

\section{Methods}

\subsection{Sample synthesis}
Pyrite-structured FeO$_{2}$ samples were synthesized from hematite ($\delta$-Fe$_{2}$O$_{3}$) powders together with ultrapure oxygen gas (O$_{2}$) at 80-90 GPa and 1800 K in DACs coupled with laser heating techniques. Details of the sample synthesis and characterization are described in Liu  \textit{et al} \cite{Liu2017}.

\subsection{High-pressure experiments}
High-pressure experiments were performed on FeO$_{2}$ up to 94 GPa using diamond-anvil cell (DAC) techniques combined with X-ray emission spectroscopy (XES), and X-ray diffraction (XRD) at beamlines 16ID-B and 16ID-D of the High Pressure Collaborative Access Team (HPCAT), Advanced Photon Source (APS), Argonne National Laboratory (ANL). 

High-pressure XES experiments were performed with an incident X-ray beam of an energy of 11.3 keV and a bandwidth of $\sim$1 eV at beamline 16ID-D, APS, ANL. Each XES spectrum was collected for approximate 1 hr and the 2 to 4 spectra at a given pressure were added together for good statistics between 30 and 91 GPa at 300 K. The pressure was determined based on the Raman spectra of the diamond anvils.

XRD experiments were conducted with a highly monochromatized incident X-ray beam of an energy of 33.17 keV (0.3738 \AA) at beamline 16ID-B, APS, ANL. The X-ray beam was focused down to 2$\sim$5 $\mu$m in the full-width at half-maximum (FWHM) at sample position.

\subsection{Calculation Details}

DFT calculations were performed by unsing WIEN2k code which uses the full-potential augmented plane wave method\cite{Blaha2001}. The Perdew-Burke-Ernzerhof (PBE) generalized gradient approximation (GGA) was used for the exchange correlation functional\cite{Perdew1997}. A 12 $\times$ 12 $\times$ 12 $k$-point mesh is used for self-consistent calculation. For structural optimization at different volumes, we used the Vienna $ab initio$ package (VASP)\cite{Kresse1996}, where a plane-wave cufoff is set to 500 eV and a 10 $\times$ 10 $\times$ 10 $k$-point mesh is used. The correlation effect of Fe 3$d$ orbitals is treated by a DMFT loop\cite{Haule2010} on the top of an effective one-electron Hamiltonian generated from the WIEN2k calcuation. The impurity model was solved by using continuous time quantum Monte Carlo (CTQMC)\cite{Haule2007}. 

\bibliography{reference}

\section{Acknowledgements}
This research was supported by a National Research Foundation of Korea (NRF) grant funded by the Korea Government (MSIP) (Grant No. 2015R1A2A1A15051540), and the Supercomputing Center/Korea Institute of Science and Technology Information with supercomputing resources including technical support (Contract No. KSC-2016-C1- 0003). D.Y.K. acknowledges financial support from the NSAF (Grant No. U1530402). W.L. Mao and J. Liu acknowledge support from the Geophysics Program by the NSF (EAR 1446969).

\section{Author Contributions}
B.G.J., D.Y.K., K.H. and J.H.S. performed the DFT and DFT+DMFT calculations. J.L., Q.H., H.-K.M. and W.L.M. conceived the high-pressure experiments. B.G.J., D.Y.K. and J.H.S. co-wrote the manuscript. All the authors discussed the results and commented on the paper.

\end{document}